# Coherent perfect absorber and laser in purely imaginary conjugate metamaterials

Yangyang Fu[1,2], Yanyan Cao[1], Steven A. Cummer[3], Yadong Xu[1*], and Huanyang Chen[1,2†]

1.College of Physics, Optoelectronics and Energy, Soochow University, No.1 Shizi Street, Suzhou 215006, China.

2.Institute of Electromagnetics and Acoustics and Department of Electronic Science, Xiamen University, Xiamen 361005, China.

3.Department of Electrical and Computer Engineering, Duke University, Durham, North Carolina 27708, USA

*Abstract:* Conjugate metamaterials, in which the permittivity and the permeability are complex conjugates of each other, possess the elements of loss and gain simultaneously. By employing a conjugate metamaterial with a purely imaginary form, we propose a mechanism for realizing both coherent perfect absorber (CPA) and laser modes, which have been widely investigated in parity-time symmetric systems. Moreover, the general conditions for obtaining CPA and laser modes, including obtaining them simultaneously, are revealed by analyzing the wave scattering properties of a slab made of purely imaginary conjugate metamaterials. Specifically, in a purely imaginary conjugate metamaterial slab with a sub-unity effective refractive index, perfect absorption can be realized for the incident wave from air.

## I. Introduction

Metamaterials [1, 2] provide unprecedented new approaches to manipulate the electromagnetic (EM) wave propagation. With them, a number of amazing optical phenomena and devices have been proposed and well demonstrated in experiments, such as the invisibility cloak [3, 4], the perfect lens [5, 6], field rotators [7, 8], and illusion optics [9, 10]. All these cases rely on control of the real parts of both permittivity and permeability of metamaterials. On the other hand, the imaginary parts (*i.e.,* the material gain or loss) also exhibit significant impact on the propagation behavior and characteristics of EM waves. For example, by constructing balanced loss and gain in materials to meet the parity-time (PT) symmetry condition, *i.e.,* $n(x) = n(-x)^*$, many significant propagation effects of light have been revealed, such as unidirectional invisibility [11, 12], coherent perfect absorption (CPA) [13-15], lasing [16, 17], and extraordinary nonlinear effects [18, 19]. Furthermore, by introducing the gain or loss acted as new freedoms to classify all possible metamaterials, the conventional two-dimensional plane could be expanded to three-dimensional space. Such operation greatly enriches the types of metamaterials, and most of their EM properties are unclear, deserving to further investigate.

One group of interest are the so-called conjugate metamaterials (CMs) [20] whose relative permittivity and permeability are complex conjugate with each other, *i.e.*, $\varepsilon' = \varepsilon e^{-i\alpha}$ and $\mu' = \mu e^{i\alpha}$, where $\varepsilon$ and $\mu$ are positive numbers and $\alpha$ is the phase factor of the materials. Although they possess the elements of loss and gain simultaneously, their refractive indexes are strictly real, and in such media unattenuated propagation of EM waves can be obtained [21, 22]. Moreover, the EM properties of CMs are largely dependent of the phase factor $\alpha$. It has been demonstrated in Ref. [20] that for $0 \leq \alpha < \pi/2$, CMs have positive refractive index; while for $\pi/2 < \alpha \leq \pi$, CMs have negative one. Such CMs might be used to serve as subwavelength-resolution lens with a perfect lens as the limiting case [20].

---

* ydxu@suda.edu.cn

† kenyon@xmu.edu.cn

Particularly, when $\alpha = \pi/2$, the corresponding CMs are purely imaginary conjugate metamaterials (PICMs), which are given by $\varepsilon' = -i\varepsilon$ and $\mu' = i\mu$. As the transition case [20] from positive to negative refractive index, PICMs offer interesting underlying physics to explore. In this work, owing to complex conjugate property, $\varepsilon = \mu$ is set for PICMs to uncover significant new behavior of waves. For simplicity, $\varepsilon = \mu = n'$ is assumed. While if they are different ($\varepsilon \neq \mu$), similar wave propagations will still happen. We thoroughly study EM wave behavior in PICMs, and find that both CPA and laser modes can be supported in a slab composed of PICMs. By analyzing the wave scattering properties of a PICM slab, we deduce the general conditions for realizing CPA and laser modes, which are verified by numerical simulations. We also derive the conditions for obtaining CPA and laser modes simultaneously. Furthermore, when the PICMs are provided with a sub-unity effective refractive index, there is critical angle for wave impinging from air to them. In contrast to the traditional case, where total reflection happens for the incident angle beyond the critical angle, a PICM slab can function as a perfect absorber beyond the critical angle, absorbing the incident wave without any reflection and transmission.

## II. Eigen modes analysis for CPA and laser in a PICM slab

We first investigate wave scattering from a PICM slab in air, which is schematically shown in Fig. 1(a). The blue and red arrows are the incoming and outgoing waves respectively, and the black arrows are the forward and backward waves inside the slab. The EM parameters of PICMs are given as $\varepsilon' = -i\varepsilon$ and $\mu' = i\mu$ ($\varepsilon = \mu = n'$) for the transverse electric (TE) polarization (the electric field is along $z$ direction). In this work, we will focus on the case of TE polarization. The similar results can be obtained for the transverse magnetic (TM) polarization (the magnetic field is along $z$ direction). Here the related parameters $\varepsilon$ and $\mu$ are real numbers, as a result the effective refractive index of PICMs is $n' = \sqrt{\varepsilon'\mu'} = \sqrt{\varepsilon\mu}$, which is a real number. The EM parameters of PICMs contain gain and loss simultaneously. Therefore, depending on which dominates in our PICM slab, an incident wave might induce quite intense transmission and reflection defined as laser modes, and as a result the injected signal (blue arrows) can be ignored (see schematic diagram in Fig. 1(b)). Or, if the loss dominates, the slab may support coherent perfect absorber (CPA) modes, which might absorb incoming coherent waves without any outgoing waves (or reflections). The schematic diagram for obtaining CPA mode is displayed in Fig. 1(c).

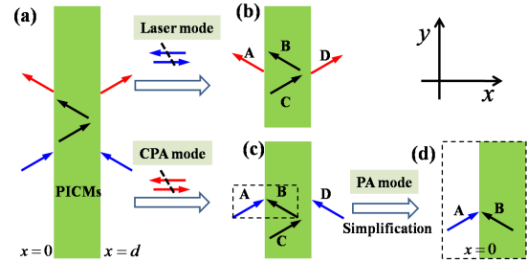

FIG. 1. (Color Online) (a) The general schematic diagram of wave scattering for the PICM slab in air. (b) and (c) The corresponding schematic diagrams of wave scattering for laser modes and coherent perfect absorber (CPA) modes, respectively. (d) The schematic diagram of wave scattering for perfect absorber (PA) modes in semi-infinite spaces consisting of air and PICM medium, which are treated as the simplified cases of CPA modes.

By analyzing the wave scattering of laser modes and CPA modes (for details, see in Appendix A), the dispersion relationships of laser modes are expressed as,

$$\eta = -i\cot(k'_x d/2) \text{ (odd modes)}, \tag{1}$$

$$\eta = i\tan(k'_x d/2) \text{ (even modes)}, \tag{2}$$

and the corresponding dispersion relationships of CPA modes are written as,

$$\eta = i\cot(k'_x d/2) \text{ (odd modes)}, \tag{3}$$

$$\eta = -i\tan(k'_x d/2) \text{ (even modes)}, \tag{4}$$

where $\eta = k_x \mu'/k'_x$, $k_x = \sqrt{k_0^2 - \beta^2}$, $k'_x = \sqrt{n'^2 k_0^2 - \beta^2}$, and $n'^2 = \varepsilon'\mu' = \varepsilon\mu$, $\beta$ is the propagation constant along $y$ direction, and $k_x$ and $k'_x$ are the wavevectors along $x$ direction in air and PICMs, respectively. Even modes are defined as symmetric modes, *i.e.*, the field distribution in $x$ direction is symmetric, and odd modes are anti-symmetric modes, *i.e.*, the field distribution in $x$ direction is anti-symmetric. In fact, for a single slab composed of PICMs ($\varepsilon = \mu$) or their analogs ($\varepsilon \neq \mu$), the above dispersion relationships are satisfied. Accordingly, CPA and laser modes can be perfectly excited owing to the real solutions for $\beta$ in Eqs. (1)-(4). While for CMs with $\alpha \neq \pi/2$, *i.e.*, CMs with real parts, there are complex solutions for $\beta$. As a result, perfect excitation for CPA and laser modes will not happen and the performance of CPA and laser modes will decrease (for more detailed discussions, please see in Appendix A). Furthermore, under some circumstances (we will explain in the following part of $0 < n' < 1$), the problem of two coherent incoming waves for CPA modes in a PICM slab could be treated as the case of a single incoming wave for perfect absorber (PA) modes in the semi-infinite spaces consisting of air and PICMs (see the schematic in Fig. 1(d), which is simplified from Fig. 1(c) by seeing the black dashed frames). The dispersion relationship for PA modes in the semi-infinite spaces is calculated as (for details, see in Appendix A),

$$\frac{k_x \mu'}{k'_x} = 1, \tag{5}$$

where $k_x = \sqrt{k_0^2 - \beta^2}$, $k'_x = i\sqrt{\beta^2 - n'^2 k_0^2}$, and $n'^2 = \varepsilon'\mu'$. From the above equations, we find that $n'=1$ is a special value, as $k_x = k'_x$ and the effective refractive index of PICMs is 1 and thus equal to that in the air background medium. Therefore, in the following we will explore the case of a PICM slab with $n' \geq 1$ in air, and then investigate the case of a PICM slab with $0 < n' < 1$ in air.

**III. CPA and laser in PICMs with $n' \geq 1$**

In this section, we will investigate the PICM slab with $n' \geq 1$. Based on Eqs. (1)-(4), the corresponding dispersion relationships between the effective refractive index $n'$ of PICM slab and the propagation constant $\beta$ are displayed in Figs. 2(a) and 2(b), which are the cases of laser modes and CPA modes respectively. The red and blue curves are the corresponding even modes and odd modes. For $n'=1.5$ as an example (see Figs. 2(a) and 2(b)), we choose the propagation constants $\beta=0.627k_0$ and $\beta=0.577k_0$ to obtain laser mode and CPA mode respectively. To verify the analytical results, we carry out numerical simulations by using COMSOL Multiphysics based finite element method. For instance, to match the tangential momentum of the laser mode, *i.e.*, $\beta=0.627k_0$, the wave with incident angle $\theta = 38.80°$ is striking from the left side on the PICM slab (see the white arrow in Fig. 2(c)), then the incoming wave oscillates inside the slab to accumulate energy, similar to the mechanism of laser modes in the cavity systems with gain media. As a result, the amplified electric field emits toward -x and +x directions by observing the field pattern and energy flux (black arrows) in Fig. 2(c), in which quite intense reflection and transmission take place in the left and right sides of the PICM slab.

To excite a CPA mode, the two incoming waves must be coherent with a specific amplitude and phase relationship at the boundaries (for details, see in Appendix A). With $A$ and $D$ as the complex amplitudes of the two incoming coherent waves, CPA requires either $A = D$ (symmetric mode) or $A = -D$ (anti-symmetric mode). When the in-phase coherent waves ( $A = D$ ) with $\theta = \pm 35.20°$ are incident from the left and right sides (see a pair of white arrows in Fig. 2(d)), they match one of the tangential wave numbers of CPA modes, i.e., $\beta=0.577 k_0$. As such, CPA is produced by the combination of destructive interference and dissipation in the PICM slab, which is well demonstrated by observing the field pattern and energy fluxes (black arrows) in Fig. 2(d). Therefore, both laser modes and CPA modes are supported in a PICM slab, if the tangential momenta are matched.

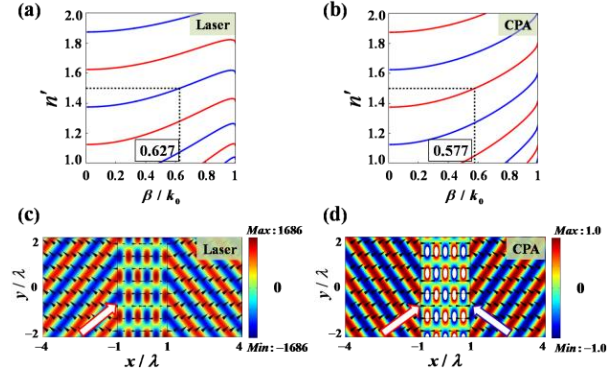

FIG. 2. (Color Online) (a) and (b) The corresponding dispersion relationships ( $n'$ vs $\beta$ ) for laser modes and CPA modes, respectively, where the red and blue curves are the corresponding even modes and odd modes. (c) and (d) The simulated electric field patterns for a laser mode and a CPA mode, respectively (for $n'$=1.5 ). In all the cases, $n' \geq 1$, $d = 2\lambda$ and $\lambda=1$.

## IV. CPA and laser in PICMs with $0 < n' < 1$

As we know, total internal reflection should occur when light is incident from a medium with a higher refractive index to a medium with a lower one. It turns out that the physics are substantially different for PICMs with lower refractive indexes. We consider a PICM slab with $0 < n' < 1$ placed in air schematically shown in Fig. 3(a). In the following, we take $\varepsilon = \mu = n' = 0.5$ for the PICM slab as an example to reveal the underlying physics. Based on the general coefficients of transmission and reflection of a PICM slab (see the Eq. (S18) and Eq. (S19) in Appendix B), the corresponding relationships between transmission/reflection of the PICM slab and the incident angle are shown in a logarithmic scale in Fig. 3(b), where the red curve is the reflection ( $R = |r|^2$ ) and the blue dashed curve is the transmission ( $T = |t|^2$ ). From Fig. 3(b), we find that when the incident angle is near $\theta = 18.00°$ or $\theta = 27.60°$, the transmission and reflection are extremely large. Moreover, when the incident angle is near $\theta$=25.60° or $\theta$=39.23° , the reflection is zero. In particular, for the incident angle with $\theta$=25.60° , the transmission is unity, which means that total transmission occurs. However, for the incident angle with $\theta = 39.23°$ , the transmission is almost zero, implying that "no transmission and no reflection" occurs.

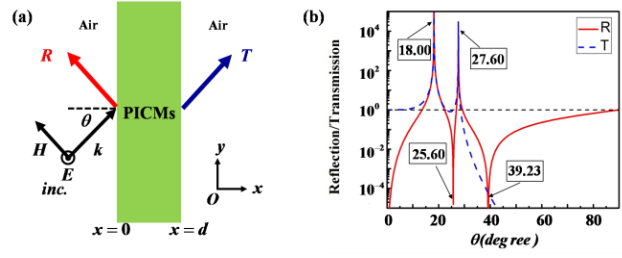

FIG. 3. (Color Online) (a) The schematic diagram of a PICM slab placed in air. (b) The relationships between transmission\reflection of the PICM slab and the incident angle in a logarithmic scale. The working wavelength is $\lambda = 1$ and the thickness of slab is $d = 2\lambda$. For the slab, we set $\varepsilon' = -0.5i$ and $\mu' = 0.5i$.

To verify the above intriguing results, numerical simulations are performed by using COMSOL Multiphysics. For example, when the wave with incident angle of $\theta = 18.00°$ impinges on the slab, the corresponding field pattern is shown in Fig. 4(a), with strong transmission and reflection. Likewise, for the wave with an incident angle of $\theta = 27.60°$, the predicted strong transmission and reflection are seen in the field pattern in Fig. 4(b). The simulated field pattern for the incident wave with $\theta = 25.60°$ is shown in Fig. 4(c), where the incident wave passes through the slab without any reflection and with unity transmission. For the incident wave with $\theta = 39.23°$, which is beyond the critical angle $\theta_c = \arcsin n' = 30°$, the corresponding field pattern is shown in Fig. 4(d), and it seems that the incident wave is bounded at the interface between air and PICM just like a surface wave, without any reflection and transmission.

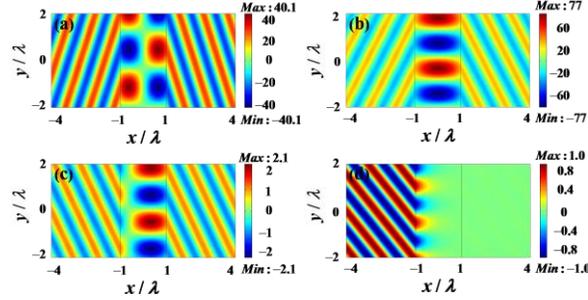

FIG. 4. (Color Online) The simulated electric field patterns for the TE plane wave incident on the PICM slab with different incident angles. (a) $\theta = 18.00°$, (b) $\theta = 27.60°$, (c) $\theta = 25.60°$, (d) $\theta = 39.23°$.

To explain the above intense and vanishing transmission and reflection, we employ Eqs.(1)-(5) to display the corresponding dispersion relationships ($n'$ vs $\beta$) for laser modes and CPA modes in Figs. 5(a) and 5(b), where the red curve and blue curve are the corresponding even modes and odd modes and the green one is related to PA modes. For the case of $n' = 0.5$ shown in Figs. 5(a) and 5(b), the corresponding tangential propagation constants marked by the black dashed lines are $0.31k_0$, $0.46k_0$, and $0.63k_0$, which are consistent with the incident angles $\theta = 18.00°$, $\theta = 27.60°$ and $\theta = 39.23°$. Therefore, the unusual transmission and reflection in Figs. 4(a) and 4(b) and Fig. 4(d) are related to the excited laser modes and CPA mode, respectively. In fact, from Figs. 5(a) and 5(b), laser and CPA modes can be excited in the PICM slab with $0 < n' < 1$, if the incident angle $\theta$ is less than $\theta_c$. However, for $\theta$ larger than $\theta_c$, the transmission will be tremendously reduced, as total internal reflection happens (*e.g.,* see the transmission curve in Fig. 3(b)). As a result, the laser modes do not exist for $\theta > \theta_c$, as wave decay will take place in the PICM slab.

However, CPA modes do exist in the case of $\theta > \theta_c$. For example, when the two incoming waves are incident on the slab, they will be bounded at the left and right interfaces without any reflection. For PICMs possessing higher refractive indices, the bounded waves inside the slab with a fixed thickness (here $d = 2\lambda$) do not interact with each other as their decay lengths are short. As a result (see Figs. 1(c) and 1(d)), the problem of two coherent incoming waves for CPA modes in a PICM slab will be treated as the case of a single incoming wave for perfect absorber (PA) modes in the semi-infinite spaces consisting of air and PICMs. Consequently, PA modes, which can perfectly absorb incident wave, could be realized in the PICM slab by replacing the coherent waves with a single incoming wave, which has been demonstrated in Fig. 4(d). This phenomena is shown clearly in the dispersion relationships in Fig. 5(b), where the lowest blue curve of CPA modes in the case of $\theta > \theta_c$ is coincident with the green curve of PA modes in the region of $n' > 0.25$, implying that CPA modes can be replaced by PA modes.

For $n' < 0.25$, marked in the gray region in Fig. 5(b), PA modes gradually deviate from CPA modes. In such case, CPA modes can be realized in the PICM slab by the incoming coherent waves. For the single incident wave shining on the PICM slab, although there is no reflection, the transmitted wave will increase for the lower refractive indices of PICMs. Therefore, PA modes are not obtained in the PICM slab in this parameter regime. This is because for PICM slab with the lower refractive indices, the bound waves, which are from the two incoming coherent waves for realizing CPA mode, interact with each other inside due to longer decayed lengths.

Furthermore, to clarify the behavior of the bounded wave inside the PICM slab, we analyze the time-averaged energy fluxes along *x* and *y* directions inside the PICM slab, *i.e.*, $\bar{S}_x = -\mathrm{Re}[H_y E_z^*]/2$ and $\bar{S}_y = \mathrm{Re}[H_x E_z^*]/2$, which can effectively reveal the energy flux distribution inside the slab. From the fields inside the PICM slab (for details, please see in the Appendix C), we find that $\bar{S}_y = 0$ in the PICM slab. In particular, $\bar{S}_x = k_x' \xi^2 \sin(k_x' x + \varphi)/(\omega \mu_0 \mu)$ is sinusoidal distribution inside the PICM slab for $\theta < \theta_c$, and $\bar{S}_x = \delta a_p^2 e^{-2\delta x}/(2\omega\mu_0\mu)$ is decayed distribution for $\theta > \theta_c$. We also display the simulated time-averaged energy fluxes shown in Figs. 5(c) and 5(d), where the incident angles are $\theta = 27.60 < \theta_c$ and $\theta = 39.23 > \theta_c$ for laser mode and PA mode, respectively. From the information in Figs. 5(c) and 5(d), inside the PICM slab, $\bar{S}_x$ exhibits the expected sinusoidal distribution for the excited laser mode and decayed distribution for PA mode, but $\bar{S}_y = 0$ in the PICM slab for both cases. The numerical results are thus consistent with the analytical results. Therefore, the bounded wave at the interface of PICM slab and air in Fig. 4(d) is absorbed, rather than a surface wave propagating along *y* direction. Furthermore, total transmission in Fig. 4(c) results from the Fabry-Pérot resonances, *i.e.*, $\phi = k_x' d = m\pi$.

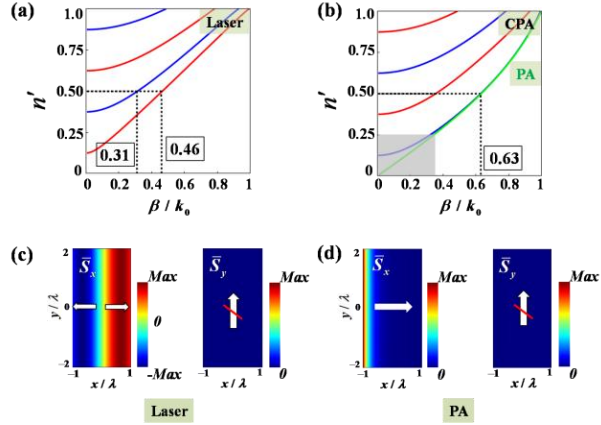

FIG. 5. (Color Online) (a) and (b) The corresponding dispersion relationships ($n'$ vs $\beta$) for laser modes and CPA modes, respectively, where the red and blue curves are the corresponding even modes and odd modes. In (b), the green curve is related to the perfect absorber (PA) modes in the semi-infinite spaces. In all the cases, $0 < n' < 1$, $d = 2\lambda$ and $\lambda=1$. (c) and (d) The simulated time-averaged energy fluxes along $x$ and $y$ directions inside the PICM slab for the cases of laser mode and PA mode, respectively.

## V. The condition for realizing CPA and laser modes simultaneously

By observing these dispersion relationships of $n' \geq 1$ (Figs. 2(a) and 2(b)) and $0 < n' < 1$ (see Figs. 5(a) and 5(b)), generally, CPA modes and laser modes take place in different conditions. In other word, for a fixed refractive index of PICM slab, CPA modes and laser modes share different propagation constants $\beta$, which have been demonstrated for the cases of $n'$=1.5 and $n'$=0.5. However, in some cases where CPA and laser modes share opposite symmetric modes, CPA modes and laser modes could be realized simultaneously. To be exact, the odd laser modes (see Eq. (1)) can have a solution identical to the even CPA modes (see Eq. (4)), or the even laser modes (see Eq. (2)) can have the same solution as the odd CPA modes (see Eq. (3)). In both cases, we can get a condition for realizing CPA modes and laser modes simultaneously with opposite symmetric modes, i.e., $\eta = k_x \mu' / k_x' = i$. After some simplification, such condition for $d = 2\lambda$ can be further given as, $\beta$=0 with $n'$=$m/2 \pm 1/8 > 0$ ($m = 1, 2, 3...$) or $n'$=1. For $\beta$=0, i.e., in the case of normal incidence [23, 24], the required effective index of the PICM is $n'$=1/8, 3/8, 5/8, ... $(2m-1)/8$, which can clearly be found in Figs. 2(a) and 2(b) and Figs. 5(a) and 5(b), where CPA modes and laser modes happen simultaneously yet with opposite symmetric modes (see the colors of dispersion curves). In addition, for $n'$=1, there are several transverse propagation constants to realize CPA modes and laser modes simultaneously, which are given as $\beta$=$0.485k_0, 0.782k_0, 0.927k_0, 0.993k_0$.

## VI. Discussion and Conclusion

In conclusion, we have analyzed the interaction of incident plane waves with uniform PICM slabs. We find, both analytically and numerically, that CPA modes and laser modes are supported in the PICM slab, and it is a mechanism different from that which occurs in PT symmetric systems, in which loss and gain are separated in different regions. In addition, the general conditions for realizing CPA and laser modes have been derived. We also identify the conditions for realizing CPA mode and laser mode simultaneously, with the case of normal incidence [22-24] as a special case. More interestingly, as a simplification of CPA modes, single-sided PA modes can be obtained in a PICM slab with a sub-unity effective index for the incident angle beyond the critical angle. Overall, we find that PICMs produce yet more unusual wave behavior that can be realized through metamaterials. We acknowledge that PICMs are very hard to realize in practice as loss and gain parameters are included in the same medium. However, recently two schemes [23, 24] to realize materials with properties similar to PICMs have been proposed. By employing PT-symmetric materials in layered structures [23], PICMs can be effectively mimicked, with the similar CPA-laser effect realized. The other method is that, based on the effective medium theory [25], PICMs could be well designed by using a photonic crystal composed of core-shell rods [24], in which loss and gain media are distributed in either the cores or the shells. Therefore, considering the recent experimental progress in optical gain [17], PICMs might be realized experimentally in the coming future.

As we focus on the CMs in this work, thus the equal amplitudes of EM parameters $\varepsilon = \mu$ are considered for PICMs. In fact, in a more general case, i.e., $\varepsilon' = -i\varepsilon$ and $\mu' = i\mu$ with $\varepsilon \neq \mu$, the similar phenomena including CPA, laser and PA can still be obtained. If we consider TM polarization in our proposed PICMs, the left sides of Eqs. (1)-(4) will change the signs as $\eta_{TM} = k_x \varepsilon' / k_x' = -k_x \mu' / k_x' = -\eta$. As a result, the dispersion relationships of laser (CPA) modes for TE polarization will be transformed into these of CPA (laser) modes for TM polarization. To be exact, for a PICM slab of interest, when it is used to realize a laser (CPA) for TE polarization, accordingly, a CPA (laser) will be achieved for TM polarization. Therefore, our proposed PICMs can be employed to realize CPA and laser modes without limit of polarizations, which is available for obtaining CPA and laser modes simultaneously. As CPA and laser modes can be effectively realized in our PICMs, PICMs also can be used to achieve negative refraction [26] and planar imaging [27], in which CPA (as the time-reversed laser [28]) and laser are respectively obtained in the loss and gain media in PT symmetric systems.

**ACKNOWLEDGMENTS**
This work was supported by the National Natural Science Foundation of China (grant No. 11604229), the National Science Foundation of China for Excellent Young Scientists (grant no. 61322504), the Postdoctoral Science Foundation of China (grant no. 2015M580456), and the Fundamental Research Funds for the Central Universities (Grant No. 20720170015). We thank Prof. Hong Chen for the helpful discussions.

**Appendix A: CPA and laser modes of PICM slab**

**Laser modes:** For the general wave scattering of a PICM slab ($\varepsilon' = -i\varepsilon$ and $\mu' = i\mu$, $\varepsilon$ and $\mu$ are positive real numbers), it includes incoming waves (the blue arrows) and outgoing waves (the red arrows), which is schematically shown in Fig. 1(a). For laser modes, the injected signals (incoming waves) can be ignored, as the outgoing waves are extremely powerful compared with the incoming waves, and the schematic of wave scattering is plotted in Fig. 1(b). Therefore, the corresponding electric field distributions in different regions can be written as,

$$\vec{E}_1 = A e^{-ik_x x} e^{i\beta y} \hat{z}, \quad x < 0, \tag{S1}$$

$$\vec{E}_2 = (B e^{ik'_x x} + C e^{-ik'_x x}) e^{i\beta y} \hat{z}, \quad 0 < x < d, \tag{S2}$$

$$\vec{E}_3 = D e^{ik_x(x-d)} e^{i\beta y} \hat{z}, \quad x > d, \tag{S3}$$

and the corresponding magnetic field distributions can be obtained by the Maxwell's Equation $\vec{H} = \nabla \times \vec{E}/i\omega\mu_0\mu$. By matching the boundary conditions at the interfaces $x=0$ and $x=d$, the dispersion relationships for laser modes are,

$$\eta = -i\cot(k'_x d/2) \text{ (odd modes)}, \tag{S4}$$

$$\eta = i\tan(k'_x d/2) \text{ (even modes)}, \tag{S5}$$

where $\eta = k_x \mu'/k'_x$, $k_x = \sqrt{k_0^2 - \beta^2}$, $k'_x = \sqrt{n'^2 k_0^2 - \beta^2}$, and $n'^2 = \varepsilon'\mu'$. Consider that the excited wave is plane wave incident from air with $\beta \leq k_0$, then $k_x = \sqrt{k_0^2 - \beta^2} > 0$. If the slab is composed of PICMs ($\varepsilon=\mu$) or their analogs ($\varepsilon \neq \mu$), $\varepsilon'$ and $\mu'$ are purely imaginary numbers:

a) For $\beta \leq n'k_0$, $\eta = k_x \mu'/k'_x$ is purely imaginary number, and the right sides of the Eq. (S4) and Eq. (S5) are also purely imaginary numbers.

b) For $\beta > n'k_0$, $\eta = k_x\mu'/k'_x$ is real number. As $k'_x = i\sqrt{n'^2k_0^2 - \beta^2}$ is purely imaginary number, the right sides of the Eq. (S4) and Eq. (S5) are real numbers.

Therefore, Eq. (S4) and Eq. (S5) can be satisfied with real solutions for $\beta$. As wave from air possesses a real tangential wavevector, perfect excitation for laser modes will happen owing to momentum matching. As a result, PICMs or their analogs can support perfect laser modes.

However, if the slab is made of CMs with real parts, i.e., $\alpha \neq \pi/2$, $\varepsilon'$ and $\mu'$ are complex numbers:

c): For $\beta \leq n'k_0$, $\eta = k_x\mu'/k'_x$ is complex number, while the right sides of the Eq. (S4) and Eq. (S5) are purely imaginary numbers.

d): For $\beta > n'k_0$, $\eta = k_x\mu'/k'_x$ is complex number; as $k'_x = i\sqrt{n'^2k_0^2 - \beta^2}$ is imaginary number, the right sides of the Eq. (S4) and Eq. (S5) are real numbers.

Therefore, Eq. (S4) and Eq. (S5) are not satisfied with real solutions for $\beta$. As wave from air possesses a real tangential wavevector, perfect excitation for laser modes will not happen due to momentum mismatching. As a result, the performance of laser modes will decrease, and CMs with $\alpha \neq \pi/2$ do not support perfect laser modes.

**Coherent perfect absorber (CPA) modes**: For obtaining CPA modes, the outgoing signals can be ignored, *i.e.,* the outgoing waves are zero, and the schematic of wave scattering is plotted in Fig. 1(c). Therefore, the corresponding electric field distributions in different regions can be written as,

$$\vec{E}_1 = Ae^{ik_xx}e^{i\beta y}\hat{z}, \quad x < 0, \tag{S6}$$

$$\vec{E}_2 = (Be^{ik'_xx} + Ce^{-ik'_xx})e^{i\beta y}\hat{z}, \quad 0 < x < d, \tag{S7}$$

$$\vec{E}_3 = De^{-ik_x(x-d)}e^{i\beta y}\hat{z}, \quad x > d, \tag{S8}$$

and the corresponding magnetic field distributions can be figured out by the Maxwell's Equation $\vec{H} = \nabla \times \vec{E}/i\omega\mu_0\mu$. By matching the boundary conditions at the interfaces $x=0$ and $x=d$, the dispersion relationships for CPA modes are,

$$\eta = i\cot(k'_xd/2) \text{ (odd modes)}, \tag{S9}$$

$$\eta = -i\tan(k'_xd/2) \text{ (even modes)}, \tag{S10}$$

where $\eta = k_x\mu'/k'_x$, $k_x = \sqrt{k_0^2 - \beta^2}$, $k'_x = \sqrt{n'^2k_0^2 - \beta^2}$, and $n'^2 = \varepsilon'\mu'$. Similar to the discussion of laser modes, perfect excitation of CPA modes can only exist in PICMs ($\varepsilon=\mu$) or their analogs ($\varepsilon \neq \mu$). For realizing CPA modes, the two incident signals should be coherent with a specific amplitude and phase relationship at the boundaries, *i.e.*, $A = (1-\eta)/(1+\eta)e^{-ik'_xd}D$.

**Perfect absorber (PA) modes:** In our PICM slab with the effective refractive index less than 1, when the incident angle $\theta$ is beyond $\theta_c$, the incident wave could be perfectly bounded at the interface between air and PICM without any reflection and transmission, defined as perfect absorber (PA) modes. When the two incoming waves for CPA are incident on the slab, they will be bounded at the left and right interfaces. For PICM slab with a fixed thickness, when PICMs possess a higher effective refractive indices, the bounded waves inside the slab are difficult to interact with each other, which can effectively throttle the outgoing waves, On this occasion, the problem of CPA modes will become the case of PA modes in the semi-infinite spaces consisting of air and PICMs as shown in Fig. 1(d). For this case, the electric field distributions in air ($x<0$) and PICM media ($x\geq 0$) can be respectively written as,

$$\vec{E}_1 = Ae^{ik_x x}e^{i\beta y}\hat{z}, \ x<0, \tag{S11}$$

$$\vec{E}_2 = Be^{-ik'_x x}e^{i\beta y}\hat{z}, \ x\geq 0, \tag{S12}$$

and the corresponding magnetic field distributions can be obtained by the Maxwell's Equation $\vec{H} = \nabla \times \vec{E}/i\omega\mu_0\mu$. By matching the boundary conditions at the interface $x=0$, the dispersion relationship for PA modes is,

$$\frac{k_x \mu'}{k'_x} = 1, \tag{S13}$$

where $k_x = \sqrt{k_0^2 - \beta^2}$, $k'_x = i\sqrt{\beta^2 - n'^2 k_0^2}$, and $n'^2 = \varepsilon'\mu'$. If the slab is composed of PICMs ($\varepsilon = \mu$) or their analogs ($\varepsilon \neq \mu$), $\varepsilon'$ and $\mu'$ are purely imaginary numbers, then Eq. (S13) can be satisfied with real solutions for $\beta$. While for CMs with real parts, there are complex solutions for $\beta$. Hence, for our proposed PICMs or their analogs, PA modes can exist.

## Appendix B: Transmission and reflection of PICM slab in air

Now we consider the reflection and refraction for a PICM slab in air (see Fig. 3(a)). Suppose the two interfaces of the CM slab and air are $x=0$ and $x=d$, where $d$ is the thickness of the slab. For an incident plane wave,

$$\vec{E}_{int} = e^{ik_x x + i\beta y}\hat{z}, \ x<0, \tag{S14}$$

the reflected wave is written as,

$$\vec{E}_{ref} = re^{-ik_x x + i\beta y}\hat{z}, \ x<0, \tag{S15}$$

while the transmitted wave is expressed as,

$$\vec{E}_{trn} = te^{ik_x(x-d) + i\beta y}\hat{z}, \ x>d, \tag{S16}$$

where $\beta = k_0 \sin\theta$ is the wave vector along the slab, $k_x = k_0 \cos\theta$ and $k'_x = k_0\sqrt{\varepsilon'\mu' - \sin^2\theta}$ are the wave vectors normal to the slab in air and CM slab, respectively, $\theta$ is the incident angle, $r$ and $t$ are the coefficients of reflection and transmission, respectively. For the PICM slab,

$$\vec{E}_{PICM} = (a_p e^{ik'_x x + i\beta y} + a_n e^{-ik'_x x + i\beta y})\hat{z}, \ 0\leq x\leq d, \tag{S17}$$

where $a_p$ and $a_n$ are the corresponding coefficients of positive and negative components. After matching the boundary conditions at the interfaces of $x=0$ and $x=d$, we have,

$$r = \frac{(1-\eta^2)2i\sin\phi}{(1+\eta)^2 \exp(-i\phi) - (1-\eta)^2 \exp(i\phi)}, \tag{S18}$$

$$t = \frac{4\eta}{(1+\eta)^2 \exp(-i\phi) - (1-\eta)^2 \exp(i\phi)}, \tag{S19}$$

and

$$a_p = \frac{2(\eta^2 + \eta)}{(1+\eta)^2 - (1-\eta)^2 \exp(2i\phi)}, \tag{S20}$$

$$a_n = \frac{2(\eta - \eta^2)\exp(2i\phi)}{(1+\eta)^2 - (1-\eta)^2 \exp(2i\phi)}, \tag{S21}$$

where $\eta = k_x \mu'/k'_x$, $\phi = k'_x d$ is the phase change in the slab.

**Appendix C: Energy flux in PICM slab**

Based on the results in Appendix B, the corresponding EM wave inside the PICM slab can be written as,

$$\vec{E}_z = (a_p e^{ik'_x x} + a_n e^{-ik'_x x})e^{i\beta y}\hat{z},  \tag{S22}$$

$$\vec{H}_x = \frac{\beta}{\omega\mu_0\mu'}(a_p e^{ik'_x x} + a_n e^{-ik'_x x})e^{i\beta y}\hat{z},  \tag{S23}$$

$$\vec{H}_y = -\frac{k'_x}{\omega\mu_0\mu'}(a_p e^{ik'_x x} - a_n e^{-ik'_x x})e^{i\beta y}\hat{z}.  \tag{S24}$$

The time-averaged energy fluxes in $x$ and $y$ directions are given as $\bar{S}_x = -\text{Re}(H_y E_z^*)/2$ and $\bar{S}_y = \text{Re}(H_x E_z^*)/2$.

a): For the EM wave in the PICM slab wave being propagating wave ($\sqrt{\varepsilon'\mu'} \geq \sin\theta$), $\beta = k_0 \sin\theta$ and $k'_x = k_0\sqrt{\varepsilon'\mu' - \sin^2\theta}$. As $\eta = k_x\mu'/k'_x$ is purely imaginary number, based on Eq. (S20) and Eq. (S21), we can easily deduce $a_p = a_n^*$, and $\vec{E}_z^* = (a_p^* e^{-ik'_x x} + a_n^* e^{ik'_x x})e^{-i\beta y}\hat{z}$, therefore,

$$\bar{S}_x = -\frac{1}{2}\text{Re}[H_y E_z^*] = -\frac{1}{2}\text{Re}[-\frac{k'_x}{\omega\mu_0\mu'}(a_p^2 e^{2ik'_x x} - a_n^2 e^{-2ik'_x x})],$$

$$\bar{S}_y = \frac{1}{2}\text{Re}[H_x E_z^*] = \frac{1}{2}\text{Re}[\frac{\beta}{\omega\mu_0\mu'}(|a_p|^2 + |a_n|^2 + a_p^2 e^{2ik'_x x} + a_n^2 e^{-2ik'_x x})].$$

We assume $a_p = \xi e^{i\varphi}$ and $a_n = \xi e^{-i\varphi}$, then

$$\bar{S}_x = -\frac{1}{2}\text{Re}[-\frac{k'_x}{\omega\mu_0\mu'}(2i\xi^2 \sin(k'_x x + \varphi))] = \frac{k'_x}{\omega\mu_0\mu}\xi^2 \sin(k'_x x + \varphi),  \tag{S25}$$

$$\bar{S}_y = -\frac{1}{2}\text{Re}[\frac{i\beta}{\omega\mu_0\mu}(2\xi^2(\cos(k'_x x + \varphi) + 1))] = 0.  \tag{S26}$$

b): For the EM waves in the PICM slab are decayed waves ($\sqrt{\varepsilon'\mu'} < \sin\theta$), $\beta = k_0 \sin\theta$ and $k'_x = ik_0\sqrt{\sin^2\theta - \varepsilon'\mu'} = i\delta$. As $\exp(2i\phi) = \exp(-2\delta d) \to 0$, we have $a_p = 2\eta/(1+\eta)$, with $\eta = k_x\mu'/k'_x = k_x\mu/\delta$, $k_x = k_0\cos\theta$, and $a_n = 0$. As a result, as $\eta = k_x\mu'/k'_x = k_x\mu/\delta$ is a real number and $a_p$ is also a real one, we have,

$$\bar{S}_x = -\frac{1}{2}\text{Re}[-\frac{i\delta}{\omega\mu_0\mu'}a_p^2 e^{-2\delta x}] = \frac{\delta}{2\omega\mu_0\mu}a_p^2 e^{-2\delta x},  \tag{S27}$$

$$\bar{S}_y = \frac{1}{2}\text{Re}[\frac{-i\beta}{\omega\mu_0\mu}a_p^2 e^{-2\delta x}] = 0.  \tag{S28}$$

Therefore, based on the above results, we can conclude that energy flux in the PICM slab only propagates along $x$ direction as $\bar{S}_y = 0$ always happens in the PICM slab.